\documentclass[journal]{IEEEtran}

\usepackage{cite}
\ifCLASSINFOpdf
\else
\fi
\usepackage{amsmath,amssymb,amsfonts}
\usepackage{algorithmic}
\usepackage{graphicx}
\usepackage{multirow}
\usepackage{makecell}
\usepackage{bm}
\usepackage{enumerate}
\usepackage{hyperref}
\def\net{SCENT }

\newcommand\toprule{\Xhline{.10em}}
\newcommand\midrule{\Xhline{.08em}}
\newcommand\bottomrule{\Xhline{.10em}}
\newcommand{\tabincell}[2]{\begin{tabular}{@{}#1@{}}#2\end{tabular}}
\begin{document}

\title{Sequence-to-Sequence Acoustic Modeling for Voice Conversion}

\author{Jing-Xuan~Zhang,
        Zhen-Hua~Ling,~\IEEEmembership{Member,~IEEE,}
        Li-Juan~Liu,
        Yuan-Jiang,
        and~Li-Rong~Dai
\thanks{This work was supported by National Key R\&D Program of China (Grant No. 2017YFB1002202), the National Nature Science Foundation of China (Grant No. 61871358) and the Key Science and Technology Project of Anhui Province (Grant No. 18030901016).}%
\thanks{J.-X. Zhang, Z.-H. Ling and L.-R. Dai are with the National Engineering Laboratory for Speech and Language Information Processing,
University of Science and Technology of China, Hefei, 230027, China (e-mail: nosisi@mail.ustc.edu.cn, zhling@ustc.edu.cn, lrdai@ustc.edu.cn). L.-J. Liu and Y. Jiang are with the iFLYTEK Co., Ltd., Hefei, 230088, China (e-mail: ljliu@iflytek.com,
yuanjiang@iflytek.com).}
\thanks{This work was conducted when J.-X. Zhang was an intern at iFLYTEK Research.}
}

\markboth{PREPRINT MANUSCRIPT OF IEEE/ACM TRANSACTIONS ON AUDIO, SPEECH AND LANGUAGE PROCESSING \copyright2018 IEEE}%
{Jing-Xuan Zhang \MakeLowercase{\textit{et al.}}: Sequence-to-sequence Acoustic Modeling for Voice Conversion}

\maketitle

\begin{abstract}
In this paper, a neural network named Sequence-to-sequence ConvErsion NeTwork (SCENT) is presented for acoustic modeling in voice conversion.
At training stage, a SCENT model is estimated by aligning the feature sequences of source and target speakers implicitly using attention mechanism.
At conversion stage, acoustic features and durations of source utterances are converted simultaneously using the unified acoustic model. 
Mel-scale spectrograms are adopted as acoustic features which contain both excitation and vocal tract descriptions of speech signals.
The bottleneck features extracted from source speech using an automatic speech recognition (ASR) model are appended as auxiliary input.
A WaveNet vocoder conditioned on Mel-spectrograms is built to reconstruct waveforms from the outputs of the SCENT model.
It is worth noting that our proposed method can achieve appropriate duration conversion which is difficult in conventional methods.
Experimental results show that our proposed method obtained better objective and subjective performance than the baseline
methods using Gaussian mixture models (GMM) and deep neural networks (DNN) as acoustic models.
This proposed method also outperformed our previous work which achieved the top rank in Voice Conversion Challenge 2018. Ablation tests further confirmed the effectiveness of several components in our proposed method.
\end{abstract}

\IEEEpeerreviewmaketitle

\section{Introduction}
\label{sec:introduction}
\IEEEPARstart{V}{oice} conversion aims to modify the speech signal of a source speaker to make it sound like being uttered by a target speaker, while keeping the linguistic contents unchanged \cite{Childers1985Voice, Childers1989Voice}.
The potential applications of this technique include entertainment, personalized text-to-speech, and so on \cite{Kain1998Spectral,Arslan1999Speaker}.

Building statistical acoustic models for feature mapping is a popular approach to  voice conversion  nowadays.
At the training stage of the conventional voice conversion pipeline, acoustic features are first extracted from the waveforms of source and target utterances.
Then, the features of parallel utterances are aligned frame by frame using alignment algorithms, such as dynamic time wrapping (DTW) \cite{2007Dynamic}.
Next, an acoustic model for conversion is trained using the acoustic features of paired source-target frames.
The acoustic model can be a joint density Gaussian mixture model (JD-GMM) \cite{Kain1998Spectral,Toda2007Voice} or a deep neural network (DNN) \cite{Desai2009voice,Desai2010Spectral}, both of which are universal function approximators \cite{Titterington1985Statistical,Hornik1989Multilayer}.
At the conversion stage, a mapping function is derived from the built acoustic model that converts the acoustic features of source speaker into those of target speaker.
Finally, waveforms are recovered from the converted acoustic features using a vocoder.

This conventional pipeline for voice conversion has its limitations. First, most previous work focused on the conversion of spectral features and simply adjusted $F_0$ trajectories linearly in the logarithm domain \cite{Desai2009voice,Desai2010Spectral, Laskar2012Comparing, Chen2014Voice, Sun2015Voice, nakashika2015voice, Lai2017Phone}.
Besides, the durations of converted utterances were kept the same as the ones of source utterances since the acoustic models were built on a frame-by-frame basis.
However, the production of human speech is a highly dynamic process and the frame-by-frame assumption constrains the modeling ability of mapping functions \cite{Mohammadi2017An}.

This paper proposes an acoustic modeling method for voice conversion based on the sequence-to-sequence neural network framework \cite{sutskever2014sequence,cho2014learning}.
A Sequence-to-sequence ConvErsion NeTwork (SCENT) is designed to directly describe the conditional probabilities of target acoustic feature sequences given source ones  without explicit frame-to-frame alignment.
The \net model follows the widely-used architecture of encoder-decoder with attention  \cite{bahdanau2015neural,luong2015effective}.
The encoder network first transforms the input feature sequences into hidden representations which are suitable for the decoder to deal with.
At each decoder time step, the attention module selects encoder outputs softly by attention probabilities and produces a context vector.
Then, the decoder  predicts output acoustic features frame by frame using context vectors.
Furthermore, a post-filtering network is designed to enhance the accuracy of the converted acoustic features.
Finally, a speaker-dependent WaveNet is utilized to recover time-domain waveforms from the predicted sequences of acoustic features.

In our proposed method, Mel-scale spectrograms are adopted as acoustic features, which do not rely on the source-filter assumption of speech production.
Therefore, $F_0$ and spectral features are converted jointly in a single model.
Additional bottleneck features derived using an automatic speech recognition (ASR) model are appended to the source Mel-spectrograms, which are expected to  improve the pronunciation correctness of the converted speech.
Attention module learns the soft alignments between the pairs of source-target feature sequence implicitly.
Facilitated by attention module, our proposed method is capable of predicting target acoustic sequences with durations different from source ones at conversion stage.

Experimental results show that our proposed method achieved better objective and subjective performance than the GMM-based and DNN-based baseline systems.
This proposed method also outperformed our previous work which achieved the top rank in Voice Conversion Challenge 2018 \cite{Liu2018}. It is worth noting that our proposed method can achieve appropriate duration conversion, which contributes to higher similarity and is difficult in conventional methods. Ablation studies were further conducted and the results confirmed the effectiveness of several key components in our proposed method.

In this paper, we focus on one-to-one voice conversion, i.e., one model is trained for each speaker pair. It should be noticed that our proposed method can also be adapted to other cases rather than one-to-one conversion. For example, the proposed method can be extended to multiple speaker pairs by conditioning on codes of speaker identities, which can be obtained from the outputs of a speaker encoder \cite{chou2018multi,arik2018neural}.

The rest of this article is organized as follows. Section~\ref{sec:relatedwork} reviews the related work on seq2seq modeling, voice cloning and WaveNet vocoders.  Section~\ref{sec:proposedmethod} introduces our proposed method for voice conversion. Details and results of experiments are presented in Section~\ref{sec:experimentsandresults}. The article is concluded in Section~\ref{sec:conclusion}.

\section{Related Work}
\label{sec:relatedwork}
\subsection{Relationship with sequence-to-sequence learning for text-to-speech}
\label{subsec:relationshipwithseq2seqtts}
Text-to-speech (TTS) methods based on seq2seq learning have emerged recently and attracted much attention \cite{wang2017tacotron:, shen2017natural,ping2018deep,article}.
Our work is inspired by the success of applying seq2seq models to TTS.
However, voice conversion is different from TTS in several aspects.
First, the inputs of a voice conversion model are frame-level acoustic features rather than phone-level or character-level linguistic features.
Typically, linguistic features are discrete, while acoustic features are continuous.
In addition to linguistic information, acoustic features also contain speaker identity information which should be processed during voice conversion.
Second, the input-output alignment in voice conversion task is different from that in TTS.
Speech generation in TTS is a \emph{decompressing} process and the alignment between text and acoustic frames is usually a \emph{one-to-many} mapping.
While the alignment can be either \emph{one-to-many} or \emph{many-to-one} in voice conversion, depending on the  characteristics of speaker pairs and the dynamic characteristics of acoustic sequences.
Third, the training data available for voice conversion  is typically smaller than that for TTS.

\subsection{Relationship with voice cloning}
\label{subsec:relationshipwithvoicecloning}

Voice cloning is a task that learns the voice of unseen speakers from a few speech samples for text-to-speech synthesis.  Unlike voice conversion, voice cloning takes text as model input.  Arik \emph{et al.} \cite{arik2018neural} evaluated two techniques of voice cloning, i.e., speaker adaptation and speaker encoding, based on Deep Voice 3 \cite{ping2018deep}.
Jia \emph{et al.} \cite{jia2018transfer} proposed a transfer learning method for voice cloning. A speaker-discriminative embedding network was first trained to achieve a speaker verification task. Then, the built network was transferred to a conditional Tacotron model \cite{wang2017tacotron:} to generate speech for a variety of speakers. Nachmani \emph{et al.} \cite{nachmani2018fitting} extended the Voice Loop model \cite{taigman2017voice} to fit new voices by incorporating a fitting network.
Instead of using text as model input in these studies, we utilize a separate ASR model for extracting linguistic-related features and the input of our model is only the speech of source speakers. Also, instead of generating speech of unseen speakers, we focus on voice conversion for one pair of speakers. It should be noticed that the techniques developed for voice cloning are potentially useful for extending our proposed method from one-to-one conversion to many-to-many conversion, which will be a part of our future study.

\subsection{Sequence-to-sequence learning for voice conversion}
To the best of our knowledge, Ramos \cite{MiguelVarelaRamos_10_2016} made the first attempt to convert spectral features using a sequence-to-sequence model with attention.
However, as stated in Section 5.5 of Ramos's thesis \cite{MiguelVarelaRamos_10_2016},
the model was not capable of using its own predictions to generate a real valued output prediction.
Kaneko \emph{et al.} \cite{Kaneko2017Sequence} proposed a CNN-based seq2seq spectral conversion method.
Because of the lack of attention module in their method, the DTW algorithm was still utilized in order to obtain frame-level aligned feature sequences during training data preparation.
Miyoshi \emph{et al.} \cite{Miyoshi2017Voice} proposed a method of mapping context posterior probabilities using seq2seq models. In their method, an RNN-based encoder-decoder converted the source posterior probability sequence to the target one for each phone, and the phone durations of  natural target speech were necessary at conversion stage.

Our work is most similar to Ramos's one \cite{MiguelVarelaRamos_10_2016}, where an utterance-level seq2seq with attention model is built for acoustic feature conversion.
Different from previous methods, Mel-spectrograms are adopted as acoustic features in our method. Thus, $F_0$ and spectral features are transformed jointly.
Our method has the ability of modeling pairs of input and output utterance without dependency on DTW alignment.
During conversion, the durations of generated target acoustic sequences are determined automatically and the probability of completion is predicted at each decoder time step.

\subsection{ Voice conversion using WaveNet}
WaveNet \cite{denoord2016wavenet}, as a neural network-based waveform generation model, has been successfully applied to TTS and voice conversion areas \cite{Kobayashi2017Statistical,Jumpei2018statistical,Liu2018}.
Studies have shown that WaveNet vocoders outperformed traditional vocoders such as WORLD \cite{Morise2016WORLD} and STRAIGHT \cite{Kawahara1999Restructuring} in terms of the quality of reconstructed speech \cite{Wang2018A, Liu2018, YangAi2018}.
Voice conversion methods using WaveNet models have also been studied in recent years. Kobayashi \emph{et al.} \cite{Kobayashi2017Statistical} proposed a GMM-based voice conversion method with WaveNet-based waveform generation.
Liu \emph{et al.} \cite{Liu2018} proposed building WaveNet vocoders for voice conversion with limited data by model adaptation.
Directly mapping source acoustic features into target speaker's waveforms using WaveNet has also been proposed \cite{Jumpei2018statistical}.

In this paper, WaveNet vocoders are used to reconstruct the waveforms of target speakers.
WaveNet vocoders accept Mel-spectrograms as input conditions and are trained in a speaker-dependent way without using the adaptation technique described in \cite{Liu2018}.

\section{Proposed Method}
\label{sec:proposedmethod}
\subsection{Overall architecture}

\begin{figure}[!t]
\centerline{\includegraphics[width=\columnwidth]{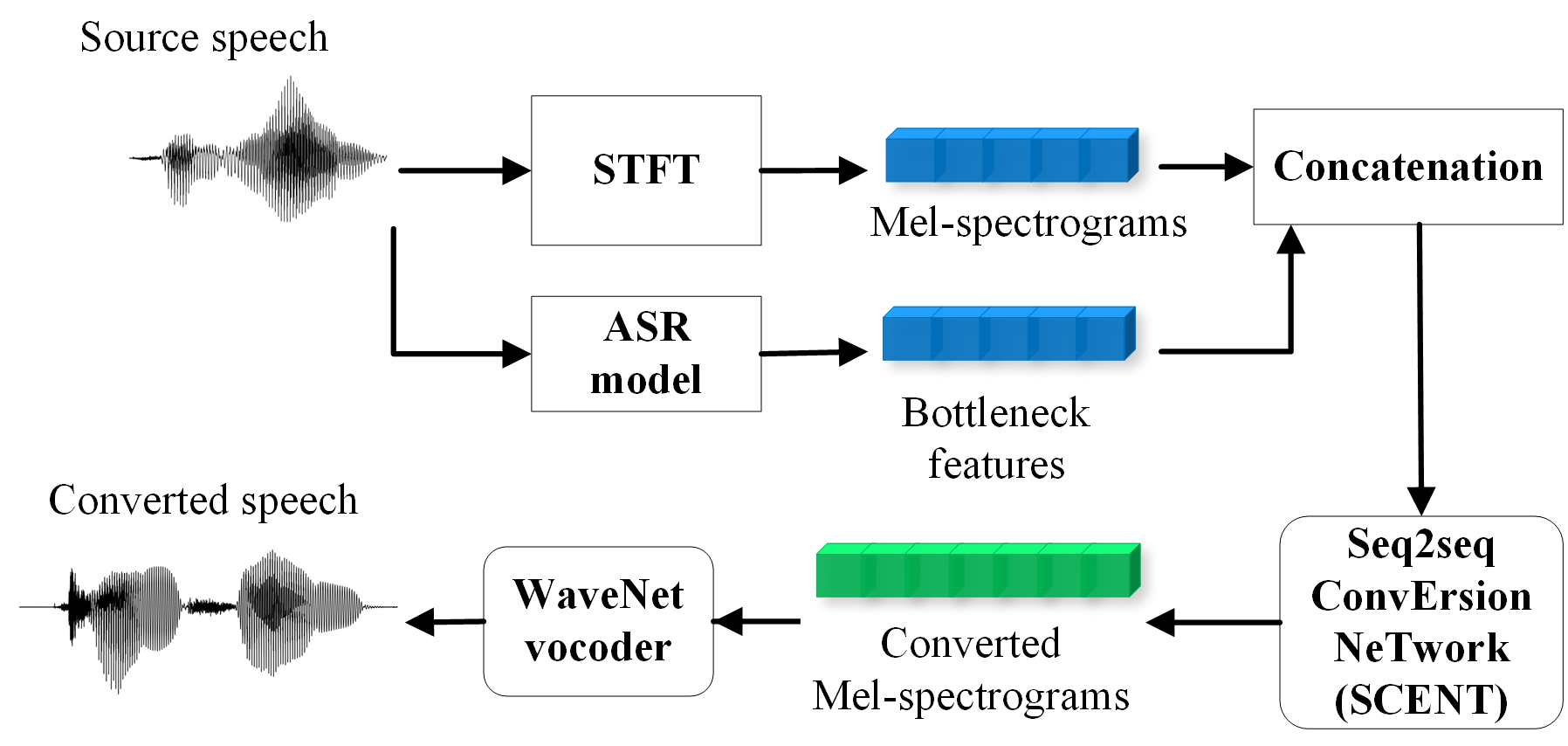}}
\caption{The conversion process of our proposed sequence-to-sequence voice conversion method.}
\label{fig1}
\end{figure}

\figurename~\ref{fig1} shows the diagram of our proposed method when converting an input utterance.
The conversion process can be divided into two main stages.
One is a Seq2seq ConvErsion NeTwork (SCENT) for acoustic feature prediction, the other is a WaveNet neural vocoder for waveform generation.
Mel-spectrograms are adopted as acoustic features in this paper.
Bottleneck features extracted by an ASR model from source speech are concatenated with acoustic features to form the input sequences of the SCENT model. The SCENT model converts input sequence into Mel-spectrograms of the target speaker.
Then, the target speaker's speech is synthesized by passing the predicted Mel-spectrograms through the WaveNet vocoder.

\subsection{Feature extraction}

Mel-spectrograms are computed through a short-time Fourier transform (STFT) on waveforms.
The STFT magnitudes are transformed to Mel-frequency scale using Mel-filterbanks followed by a logarithmic dynamic range compression.
In order to extract bottleneck features, a recurrent neural network (RNN) based ASR model is trained on a separate speech recognition dataset.
For each input frame, bottleneck features, i.e., the activations of the last hidden layer before the softmax output layer of the ASR model, are extracted.
Such bottleneck features can provide additional linguistic-related descriptions which are expected to benefit the conversion process.
It should be noticed that these bottleneck features are still automatically extracted from the acoustic signals of source utterances  and no text transcriptions are necessary.
The Mel-spectrograms and bottleneck features at each frame are concatenated to form the input sequence $\bm{x} = [\bm{x}_{1},\dots, \bm{x}_{T_x}]$ of the SCENT model,
where $T_x$ is the frame number of source speech.

\subsection{Structure of \net}
\label{subsec:structure}
\begin{figure}[!t]
\centerline{\includegraphics[width=220pt]{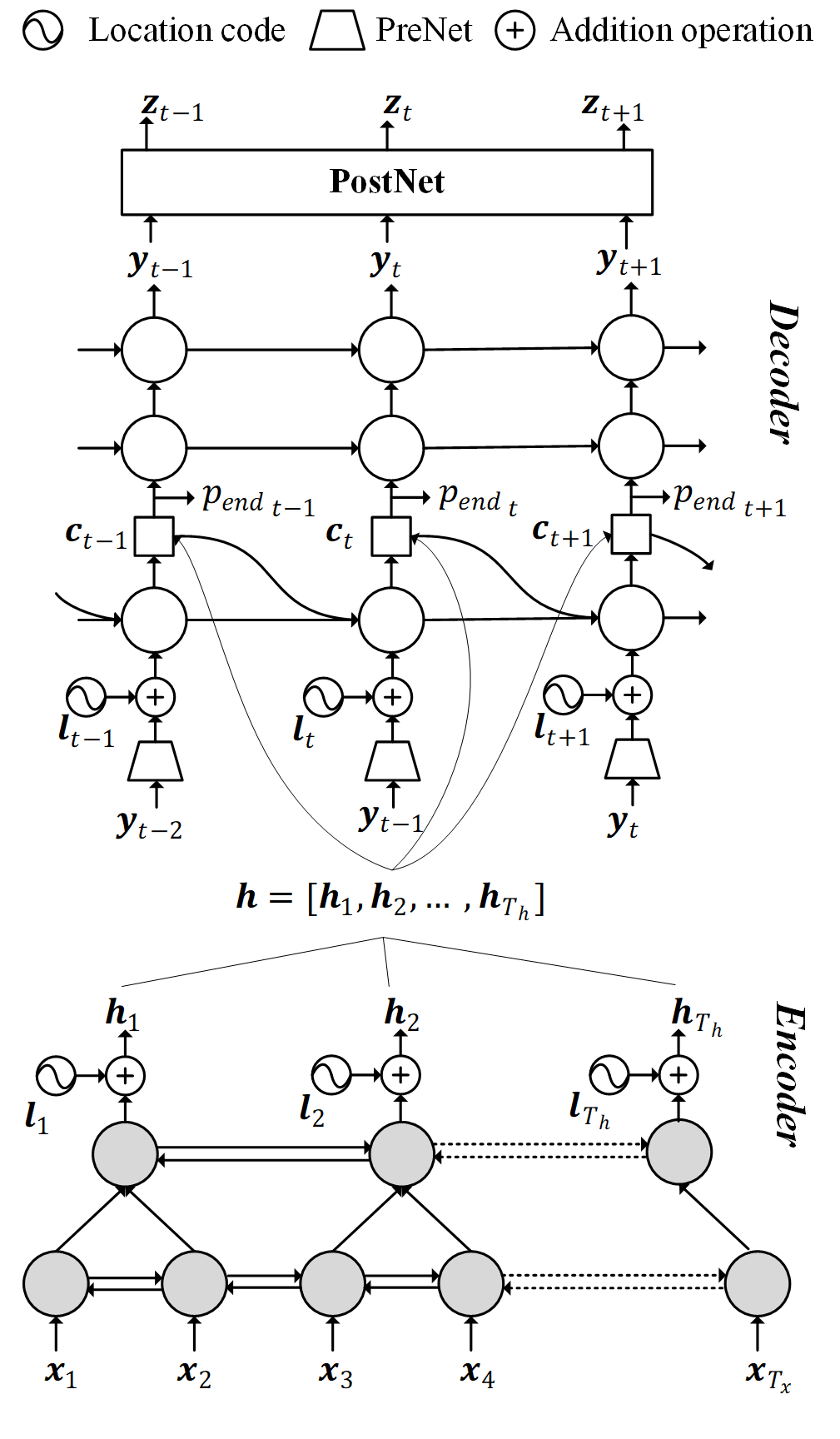}}
\caption{The network structure of a SCENT model, where skip connections and residual connections are ignored for clarity. The grey circles in the encoder represent LSTM units with layer normalization. $T_x$ and $T_h$ are the frame numbers of input sequence and hidden representations.
 The encoder in this figure  has a downsampling rate $M=2$. Therefore, we have $T_x=2T_h$ in this figure. The auto-regressive inputs of the decoder are natural history contexts at training time and are generated ones at conversion time.
 Single frame is predicted at each decoder time step (i.e., $r=1$) in this figure.}
\label{fig2}
\end{figure}

A \net model contains an encoder-decoder with attention network which predicts acoustic feature in an uni-directional left-to-right way, and a bi-directional post-filtering network which refines the generation results.
\figurename~\ref{fig2} shows the network structure of a SCENT model.

Let $\bm{y} = [\bm{y}_{1},\dots, \bm{y}_{T_y}]$ denote the output Mel-spectrogram sequence of the encoder-decoder network, where $T_y$ is the frame number of  target speech.
The encoder-decoder network models the mapping relationship between input and output feature sequences using conditional distributions of each output frame $\bm{y}_{t}$ given previous output frames $\bm{y}_{<t}=[\bm{y}_1,\dots,\bm{y}_{t-1}]$ and the input $\bm{x}$ as
\begin{equation}
p(\bm{y}|\bm{x})=\prod_{t=1}^{T_y}{p(\bm{y}_{t}|\bm{y}_{<t},\bm{x},W_{enc},W_{dec})},
\label{eq1}
\end{equation}
where $W_{enc}$ and $W_{dec}$ are parameters of the encoder-decoder network.
As shown in \figurename~\ref{fig2}, the encoder transforms the concatenated Mel-spectrograms and bottleneck features of source speech into a high-level and abstract representation $\bm{h}=[\bm{h}_1,\dots,\bm{h}_{T_h}]$ as
\begin{equation}
\bm{h}=\text{Encoder}(\bm{x}, W_{enc}).
\label{eq2}
\end{equation}
$T_h$ is the frame number of hidden representations and $T_h<T_x$ because of the pyramid structure of encoder.
The decoder with attention mechanism utilizes $\bm{h}$ and produces a probability distribution over output frames as
\begin{equation}
p(\bm{y}_t|\bm{y}_{<t},\bm{x})=\text{Decoder}(\bm{y}_{<t},\bm{h},W_{dec}).
\label{eq3}
\end{equation}

The generation process of the decoder network is uni-directional. In order to make use the bi-directional context information, a post-filtering network (i.e., PostNet) is further employed to enhance the accuracy of prediction.
Let $\bm{z}=[\bm{z}_1, \dots, \bm{z}_{T_z}]$ represent the PostNet output sequence, which is the final prediction of the SCENT model.
In this paper, the frame rates of decoder outputs and PostNet outputs are the same, i.e. $T_z=T_y$. The distribution of feature sequence $\bm{z}$ given the output of the encoder-decoder network $\bm{y}$ is modeled as
\begin{equation}
p(\bm{z}|\bm{y})=\text{PostNet}(\bm{y},W_{pos}),
\label{eq4}
\end{equation}
where $W_{pos}$ denotes the parameters of the PostNet.

Next, we will describe each part of \net in details.

\subsubsection{Encoder}
The encoder network is constructed based on the pyramid bidirectional LSTM architecture \cite{hochreiter1997long, Chan2016Listen}, which processes the sequence with lower time resolution at higher layers.
In a conventional deep bidirectional LSTM (BLSTM) architecture, the output at the $n$-th time step of the $j$-th layer is computed as
\begin{equation}
\bm{h}_{n}^{j}=\text{BLSTM}(\bm{h}_{n-1}^{j}, \bm{h}_{n}^{j-1}).
\label{eq5}
\end{equation}
In a pyramid BLSTM (pBLSTM), the outputs at consecutive steps of a lower layer are concatenated and fed into the next layer to decrease the sampling rate of input sequence.
The general calculation of pBLSTM hidden units can be formulated as
\begin{equation}
\bm{h}_{n}^{j}=\text{pBLSTM}(\bm{h}_{n-1}^{j}, [\bm{h}_{M*n}^{j-1},\dots,\bm{h}_{M*n+M-1}^{j-1}]),
\label{eq6}
\end{equation}
where $M$ is ratio of downsampling.
The technique of layer normalization \cite{2016arXiv160706450L} is applied to the encoder LSTM cells.
Then, a location code $\bm{l}_n=[l_n(0),\dots,l_n(d-1)]^\top$ \cite{vaswani2017attention} is added to the top output layer of pBLSTMs to form the hidden representation $\bm{h}$.
Let $d$ be the dimension of each $\bm{h}_n$. The location code is composed of sine and cosine functions of different frequencies as
\begin{equation}
\left\{
    \begin{array}{lr}
    l_{n}(2i) =  \sin(\left.n \middle / 10000^{ \left.2i\middle/d\right.} \right.), \\
    l_{n}(2i+1) = \cos(\left.n\middle/10000^{\left.2i\middle/d\right.}\right.),
    \end{array}
\right.
\label{eq7}
\end{equation}
where $n$ is the time step in sequence $\bm{h}$ and $i\in[0,\dots,\left.d\middle/2\right.-1]$ is the dimension index.
The base 10000 in Eq.~(\ref{eq7}) follows the configuration in the original paper \cite{vaswani2017attention} which proposed the location code.
This location code is useful since it gives the model explicit information of which portion of the sequence is currently processed. The effectiveness of the location code will be demonstrated by ablation tests in our experiments.

The pyramid structure of our encoder network results in shorter hidden representation than original input sequence.
For the voice conversion task, we expect that the encoder network should exclude speaker-dependent information of the source speech and extract hidden representation $\bm{h}$ which is high-level and linguistic-related.
Because one phone usually corresponds to tens of acoustic frames, it is reasonable to derive hidden representation with lower sampling rate than the frame-level input sequence.
Furthermore, hidden representation with lower sampling rate makes the attention module easier to converge, since this leads to fewer encoding states for attention calculation at each decoding step.
This pyramid structure also reduces the computational complexity by shortening the length of $\bm{h}$ for attention calculation and speeds up training and inference significantly.

\subsubsection{Decoder with attention mechanism}
The decoder is an auto-regressive RNN which predicts the output acoustic features from the hidden representation $\bm{h}$.
Non-overlapping $r$ frames are predicted at each decoder step. This trick divides the total decoding steps by $r$, which further reduces training and inference time \cite{wang2017tacotron:}.
In \figurename~\ref{fig2}, the decoder is illustrated with $r=1$ for clarity.
The prediction of previous time step $\bm{y}_{t-1}$ is first passed through a pre-processing network (i.e., PreNet), which is a two-layer MLP with ReLU activation and dropout in our implementation.
The MLP outputs are sent into an LSTM layer with attention mechanism.
A context vector $\bm{c}$ is  calculated at each decoder step using attention probabilities as
\begin{equation}
\bm{c}_t=\sum_{n=1}^{T_h}{\alpha_{t}^{n}\bm{h}_n},
\label{eq8}
\end{equation}
where $\bm{\alpha}_{t}=[\alpha_{t}^1,...,\alpha_{t}^{T_h}]$ are  attention probabilities, $t$ is decoder time step,
and $n$ is the index of encoder outputs.

In our implementation, a hybrid attention mechanism is adopted which takes the alignment of previous decoder step (i.e.,
location-awareness) into account when computing the attention probabilities.
In order to extract location information, $k$ filters with kernel size $l$ are employed to convolve the alignment of previous time step.
Let $F\in \mathbb{R}^{k\times l}$ represent the convolution matrix, and $\bm{q}$ denote the query vector which is given by the output of attention LSTM.
Then, the attention score $e_t^n$  is computed as
\begin{equation}
\bm{f}_t=F*\bm{\alpha}_{t-1},
\label{eq9}
\end{equation}
\begin{equation}
e_{t}^{n}=\bm{q}_t^{\top}W\bm{h}_n+\bm{v}^\top\text{tanh}(U\bm{f}_t^n + \bm{b}),
\label{eq10}
\end{equation}
where $\bm{v}$, $\bm{b}$, $W$ and $U$ are trainable parameters of the model.
As we can see from Eq.~(\ref{eq10}), the calculation of the hybrid attention takes two parts into consideration. The first part of Eq.~(\ref{eq10}) measures the relationship between the query vector and different entries of encoder outputs. 
The second part of Eq.~(\ref{eq10}) is computed based on the alignment of previous decoder step $\bm{\alpha}_{t-1}$
and provides a constraint on current attention probabilities. 
The convolution matrix is employed to filter $\bm{\alpha}_{t-1}$ for extracting useful features as shown in Eq.~(\ref{eq9}).
The features are further integrated into the calculation of attention scores as shown in Eq.~(\ref{eq10}). 

Furthermore, the forward attention method proposed in our previous work \cite{forward} is adopted to stabilize the attention alignment and speed up the convergence of attention alignment.
In the forward attention method, the attention probability $\alpha_t^n$ is calculated
as
\begin{equation}
\hat{e}_{t}^{n}=\left.\exp(e_t^n)\middle/\sum_{i=1}^{T_h}{\exp(e_t^i)}\right.,
\label{eq11}
\end{equation}
\begin{equation}
\hat{\alpha}_t^n = \hat{e}_t^{n}(\alpha_{t-1}^n+\alpha_{t-1}^{n-1}),
\label{eq12}
\end{equation}
\begin{equation}
\alpha_t^n = \left.\hat{\alpha}_t^n\middle/\sum_{i=1}^{T_h}{\hat{\alpha}_t^i}\right..
\label{eq13}
\end{equation}
For initialization, we have
\begin{equation}
\left\{
    \begin{array}{lr}
    \alpha_0^1 = 1, & \\
    \alpha_0^n = 0, & \text{for } n = 2,\dots, T_h.
    \end{array}
\right.
\label{eq14}
\end{equation}

The motivation of forward attention is to follow the monotonic nature of alignments in human speech generation \cite{forward}. Therefore, a forward variable which only takes the monotonic alignment paths into consideration is designed. This forward variable is derived from the original attention probabilities $\hat{e}_{t}^{n}$ and it can be computed recursively as Eq.~(\ref{eq12}). Then, the normalized forward variables $\alpha_t^n$ are used to replace original attention probabilities $\hat{e}_{t}^{n}$ for summarizing the encoder outputs as shown in Eq.~(\ref{eq8}).
In addition, a location code is also added to the auto-regressive input of the decoder at each time step.

The context vector $\bm{c}$ and query vector $\bm{q}$ are concatenated and fed into a stack of two-layer decoding LSTMs.
The concatenation of $\bm{c}$, $\bm{q}$ and the outputs of decoding LSTMs are linearly projected
to produce the Mel-spectrogram output of the decoder network.
In parallel, the concatenation of $\bm{c}$ and $\bm{q}$ are linearly projected  to a scalar and passed through a sigmoid activation to predict the completion probability $p_{end}$, which indicates whether the converted sequence reaches the last frame.

\subsubsection{Post-filtering network}
The PostNet refines the Mel-spectrograms predicted by the decoder using bi-directional context information.
The PostNet is a convolutional neural network (CNN) with a residual connection from network input to the final output.
The first layer of the PostNet is composed of 1-D convolution filter banks
in order to extract rich context information. The outputs of the convolution banks are stacked together and further passed through a two-layer 1-D CNN.
The outputs of the final layer  are added to the input Mel-spectrograms to produce the final results.

\subsection{Loss function of \net}
\label{subsec:lossfunction}
We train the \net model by multi-task learning and the total loss is the weighted sum of three sub-losses as
\begin{equation}
L = w_{dec}L_{dec} + w_{post}L_{post} + w_{end}L_{end},
\label{eq15}
\end{equation}
where $w_{dec}$, $w_{post}$ and $w_{end}$ are the weights of the three components.
$L_{dec}$ and $L_{post}$ denote the losses of Mel-spectrogram prediction given by the decoder and the PostNet respectively.
$L_{end}$ is a binary cross-entropy loss for evaluating the predicted completion probabilities.

Two types of criteria are investigated for $L_{dec}$.
One is the minimum square error (MSE) between the predicted and ground truth acoustic features.
The other is the maximum likelihood (ML) criterion based on Gaussian mixture model (GMM).
For GMM-ML, the network outputs are adopted to parameterize a GMM following the framework of mixture density networks (MDN) \cite{bishop1994mixture,6854321}.

More specifically, the likelihood function in GMM-ML takes the form of a GMM  as
\begin{equation}
p(\bm{y}|\bm{x},W_{enc}, W_{dec})=\sum_{i=1}^{m}{w_i(\bm{x})\mathcal{N}\left(\bm{y};\bm{\mu}_i(\bm{x}), \bm{\Sigma}_i(\bm{x})\right)},
\label{eq16}
\end{equation}
where $m$ is the number of mixture components, and $w_i(\bm{x})$, $\mu_i{(\bm{x})}$ and $\bm{\Sigma}_i(\bm{x})$ correspond to the mixture weight, mean vector and covariance matrix of the $i$-th Gaussian component given $\bm{x}$. Here, the covariance matrices are set to be diagonal.
The concatenation of $\bm{c}_t$, $\bm{q}_t$ and the outputs of decoding LSTMs are projected to a vector $\bm{o}(\bm{x},W_{enc},W_{dec})\in \mathbb{R}^{(2d_{Mel}+1)m}$, where $d_{Mel}$ is the dimension of Mel-spectrograms and the whole vector can be divided into all mixture components as
\begin{equation}
\begin{split}
\bm{o}(\bm{x},W_{enc},W_{dec})= &[o_1^{(w)}(\bm{x}), \dots, o_m^{(w)}(\bm{x}), \\
&\bm{o}_1^{(\sigma)}(\bm{x})^{\top}, \dots, \bm{o}_m^{(\sigma)}(\bm{x})^{\top}, \\
&\bm{o}_1^{(\mu)}(\bm{x})^{\top}, \dots, \bm{o}_m^{(\mu)}(\bm{x})^{\top}]^\top.
\end{split}
\label{eq20}
\end{equation}
Then, the GMM parameters in Eq.~(\ref{eq16}) can be derived from the vector $\bm{o}(\bm{x},W_{enc},W_{dec})$ as
\begin{equation}
w_i(\bm{x})=\left.\exp{\left(o_i^{(w)}(\bm{x})\right)}\middle/{\sum_{j=1}^m{\exp{\left(o_j^{(w)}(\bm{x})\right)}}}\right.,
\label{eq17}
\end{equation}
\begin{equation}
\bm{\sigma}_i(\bm{x})=\log{\left(\exp{(\bm{o}_i^{(\sigma)}(\bm{x}))}+1\right)},
\label{eq18}
\end{equation}
\begin{equation}
\bm{\mu}_i(\bm{x})=\bm{o}_i^{(\mu)}(\bm{x}),
\label{eq19}
\end{equation}
where $\bm{\sigma}_i(\bm{x})$ is a vector composed of the diagonal elements of $\bm{\Sigma}_i(\bm{x})$.
For GMM-ML, $L_{dec}$ is defined as the negative log-likelihood (NLL) function, i.e.,
\begin{equation}
L_{dec}=-\log{ p(\bm{y}|\bm{x},W_{enc},W_{dec})}.
\label{eq21}
\end{equation}

Under both MSE and GMM-ML criteria, natural acoustic histories of target speech are sent into the decoder at training time.
The MSE criterion is actually a special case of GMM-ML which uses single mixture with fixed unit variance and predicted mean vector \cite{christopher2016pattern}.
Theoretically, GMM-ML is more flexible since it models more general probability distributions and
the MSE criterion usually leads to over-smoothed prediction because of the averaging effect \cite{bishop1994mixture}.

When applying the GMM-ML criterion to $L_{dec}$, the mean vector of the component with maximum prior probability is used to generate the output sample at both training and testing stages.
At training time, the gradients from the PostNet are only back-propagated through the sampled mean vectors given by the decoder output layer.

Only the MSE criterion is applied to $L_{post}$ in our implementation.
For calculating $L_{end}$, only the last decoder step of a natural target sequence is labelled as 1 (i.e., completed) and the rest steps are labelled as 0 (i.e., incompleted).

\subsection{WaveNet-based vocoder}
As shown in \figurename~\ref{fig1}, a WaveNet-based vocoder is adopted to reconstruct time-domain waveforms given the predicted Mel-spectrogram features.

In our WaveNet model, the Mel-spectrogram features are first passed through a ConditionNet consisting of stack of dilated 1-D convolution layers with parametric ReLU activation (PReLU) \cite{xu2015empirical}.
The outputs of ConditionNet are upsampled to be consistent with the sampling rate of waveforms by simply repeating.
Then, the sequence of condition vectors are fed into each dilated convolution block of the WaveNet to control the waveform generation.
Our WaveNet model is trained only using the target speech data for building the SCENT model and the adaptation technique \cite{Liu2018} is not used in this paper.

\section{Experiments}
\label{sec:experimentsandresults}
\subsection{Experimental conditions}
\label{expcond}

\begin{table}
\renewcommand\arraystretch{1.2}
\caption{Details of model configurations.}
\label{tab1}
\begin{tabular}{p{30pt}|p{45pt}<{\centering}|p{140pt}}
\toprule
\multirow{12}{30pt}{\net} & \multirow{2}{45pt}{Encoder}   & pBLSTM, 2 layers and 256 cells LSTM  \\
                         &                               & with layer normalization, $M=4$ \\
                         \cline{2-3}
                         & \multirow{2}{45pt}{PreNet}    &FC-256-ReLU-Dropout(0.5)$\to$ \\
                         &                               &FC-256-ReLU-Dropout(0.5) \\
                         \cline{2-3}
                         & \multirow{4}{45pt}{Decoder}  & Attention LSTM, 1 layer and 256 cells; \\
                         &                                             & $k=10$ and $l=32$ for $F$ in Eq.~(\ref{eq9}); \\
                         &                                             & $\bm{v}$ in Eq.~(\ref{eq10}) has dimension of 256; \\
                         &                                             & Decoder LSTM, 2 layers and 256 cells \\
                         \cline{2-3}
                         & \multirow{4}{45pt}{PostNet}   & Conv1D banks, $k=[1,\dots,8]$, \\
                         &                               & Conv1D-$k$-256-BN-ReLU-Dropout(0.2)$\to$ \\
                         &                               & Conv1D-$3$-256-BN-ReLU-Dropout(0.2)$\to$ \\
                         &                               &Conv1D-$3$-256-BN-ReLU-Dropout(0.2) \\
\midrule
\multirow{5}{30pt}{WaveNet
vocoder}  & \multirow{2}{45pt}{ConditionNet}   & 4 layers Conv1D-$3$-100-PReLU  \\
          &                                    & with dilation $d=[1,2,4,8]$\\
          \cline{2-3}
          & \multirow{3}{45pt}{WaveNet}        & 30 layers dilated convolution layers \\
          &                                    & with dilation $d=2^{k\bmod10}$ for \\
          &                                    & $k=[0,\dots,29]$;  1024 softmax output \\
\bottomrule
\multicolumn{3}{p{230pt}}{FC represents fully connected. BN represents for batch normalization.
Conv1D-$k$-$n$ represents 1-D convolution with kernel size $k$ and channel $n$.}
\end{tabular}
\end{table}

Two datasets were used in our experiments. The first one contained 1060  parallel Mandarin Chinese utterances from one male speaker (about 53 mins) and one female speaker (about 72 mins).
This dataset was separated into a training set with 1000 utterances, a validation set with 30 utterances and a test set with 30 utterances. For the second dataset, speech data of
one male (rms, about 62 mins) and one female (slt, about 52 mins) from the CMU ARCTIC database \cite{arctic} was adopted. This dataset contained 1132 parallel English utterances, which were separated into a training set with 1000 utterances, a validation set with 66 utterances and a test set with 66  utterances. Our analytical experiments in Section~\ref{subsec:losseva} and Section~\ref{subsec:ablation} only adopted the Mandarin dataset, and the main objective and subjective evaluations in Section~\ref{subsec:compari} adopted both datasets.

The recordings of both dataset were sampled at 16kHz.
80-dimensional Mel-scale spectrograms were extracted every 10 ms with Hann windowing of 50 ms frame length and 1024-point Fourier transform.
512-dimensional bottleneck features were extracted using an ASR model every 40 ms and  were then upsampled by repeating to match the frame rate of Mel-spectrograms.

The speaker-independent ASR model was trained using internal datasets of iFLYTEK company, which contained recordings of about 10,000 hours for Mandarin and recordings of about 3,000 hours for English. 
Our ASR model was an LSTM-HMM-based one. The LSTM  was bidirectional with 6 hidden layers and 1024 units in each direction.
 The classification targets of the LSTM model were clustered triphones, i.e., senones. For the Mandarin dataset, the phoneme  set 
included 26 initials and 140 tonal finals. 
We evaluated the performance of the ASR model on the parallel dataset for voice conversion. The frame classification accuracies for the female and male speakers were 72.3\% and 78.4\% respectively. For the English dataset, there were 62 phonemes and the frame classification accuracies for the female and male speakers were 76.4\% and 75.9\% respectively.

The details of our model configurations are listed in \tablename~\ref{tab1}.
In our implementation, two frames were predicted at one decoding step (i.e., $r=2$) and only the last frame was fed back into the PreNet for the generation at next step.
In the loss function for training the SCENT model, $w_{dec}$ was heuristically set as 1.0 or 0.01 if  MSE or GMM-ML training criterion was adopted for $L_{dec}$.
$w_{post}$ and $w_{end}$ were heuristically set as 1.0 and 0.005 respectively.
Zoneout \cite{krueger2017zoneout} with probability of 0.2  were used at LSTM layers for regularization.
Residual connections were adopted for the LSTM layers of encoder and decoder to speed up model convergence.
We used Adam \cite{kingma2014} optimizer with learning rate of $10^{-3}$ for the first 20 epochs.
After 50 epochs, the learning rate was exponentially decayed by 0.95 for each epoch. $L_2$ regularization with weight $10^{-6}$ was also applied.
The batch size was 4.
For WaveNet training, the $\mu$-law companded waveforms were quantized into 10 bits, i.e., 1024 levels.
A speaker-dependent WaveNet vocoder was trained using each speaker's waveforms with random initialization and a learning rate of $10^{-4}$ until the loss converge.

Three kind of baseline methods were adopted for comparison in our experiments.
41-dimensional Mel-cepstral coefficients (MCCs), 1-dimensional fundamental frequency ($F_0$) and 5-dimensional band aperiodicities (BAPs) were extracted every 5 ms by STRAIGHT \cite{Kawahara1999Restructuring} as acoustic features in our baseline systems.
The descriptions of these methods are as follows\footnote{Samples of audio are available at \url{https://jxzhanggg.github.io/Seq2SeqVC}.}.
\begin{itemize}
\item \textbf{JD-GMM:}
Gaussian mixture models with full-covariance matrices were utilized for modeling the joint spectral feature vectors of source and target speakers.
For each speaker, static and delta spectral features were used.
The number of mixtures $m$ was tuned on validation set with $m\in[16, 32, 48, 64]$.
Maximum likelihood parameters generation (MLPG) with global variance (GV) enhancement were used for spectral parameter generation.
$F_0$ was converted by Gaussian normalization in the logarithm domain \cite{chappell1998speaker-specific}.
BAPs were not converted but directly copied from the source, since previous research showed that converting aperiodic component did not make a statistically significant difference to the quality of converted speech \cite{Ohtani06maximumlikelihood}.
Waveforms were reconstructed by STRAIGHT vocoder from the converted acoustic features.
\item \textbf{DNN:} The DNN-based voice conversion models were implemented based on Merlin toolkit \cite{Zhizheng2016Merlin}.
The static, delta and acceleration components of MCCs, $F_0$ and BAPs were transformed jointly using a DNN.
In addition to use the acoustic features of the source speaker as model input,
we also  concatenated the input acoustic features with the bottleneck features used in our proposed method.
This approach was named \textbf{bn-DNN} in the rest of this paper.
The ReLU activation function was used at DNN hidden units.
A grid search using validation set was adopted in order to pick up the optimal depth $d$ and width $w$ of the DNN with $d\in[3, 4, 5, 6]$ and $w\in[512, 1024, 2048]$.
MLPG and GV techniques were used for acoustic parameter generation.  Waveform was reconstructed by STRAIGHT vocoder from the converted acoustic features.
\item \textbf{VCC2018:}
This baseline method followed the framework of our previous work \cite{Liu2018}, which 
achieved the top rank on naturalness and similarity in Voice Conversion Challenge 2018.
A speaker-dependent acoustic feature predictor was trained by adapting a pre-trained speaker-independent model using the data of the target speaker.
The predictor was an LSTM model which predicted MCCs, $F_0$ and BAPs of the target speaker from bottleneck features frame-by-frame.
At the training stage, bottleneck features were extracted from the target speaker as model inputs. At the conversion stage, bottleneck features were obtained from the speech of the source speaker and were sent into the acoustic feature predictor of the target speaker for conversion.
In this method, a speaker-dependent WaveNet vocoder conditioned on MCCs, $F_0$ and BAPs features was built for waveform reconstruction.
\end{itemize}

\begin{table}[!t]
\renewcommand\arraystretch{1.2}
\centering
\caption{Objective evaluation results of using different loss functions for the decoder on validation set.}
\label{tab2}
\begin{tabular}{p{20pt}|p{40pt}<{\centering}|p{40pt}<{\centering}|p{40pt}<{\centering}|p{40pt}<{\centering}}
\toprule
\multirow{3}*{Settings}&
\multicolumn{2}{c|}{Female-to-Male}&
\multicolumn{2}{c}{Male-to-Female}\\
\cline{2-5}
&MCD  &$F_0$ RMSE &MCD &$F_0$ RMSE \\
&(dB) &(Hz) &(dB) &(Hz)\\
\midrule
MSE &3.397 &42.122 &3.658 &33.420 \\
\hline
MX2 &\textbf{3.365} &\textbf{38.123} &3.649 &\textbf{32.271}\\
\hline
MX4 &3.384 &38.629 &3.651 &34.748 \\
\hline
MX6 &3.376 &38.804 &3.669 &35.337 \\
\hline
MX8 &3.418 &39.230 &\textbf{3.637} &33.029 \\
\bottomrule
\multicolumn{5}{p{230pt}}{``MX2'', ``MX4'', ``MX6'' and ``MX8'' represent using ML criterion with  2, 4, 6 and 8 GMM mixture components respectively.}
\end{tabular}

\end{table}

\begin{figure}[!t]
\centerline{\includegraphics[width=0.9\columnwidth]{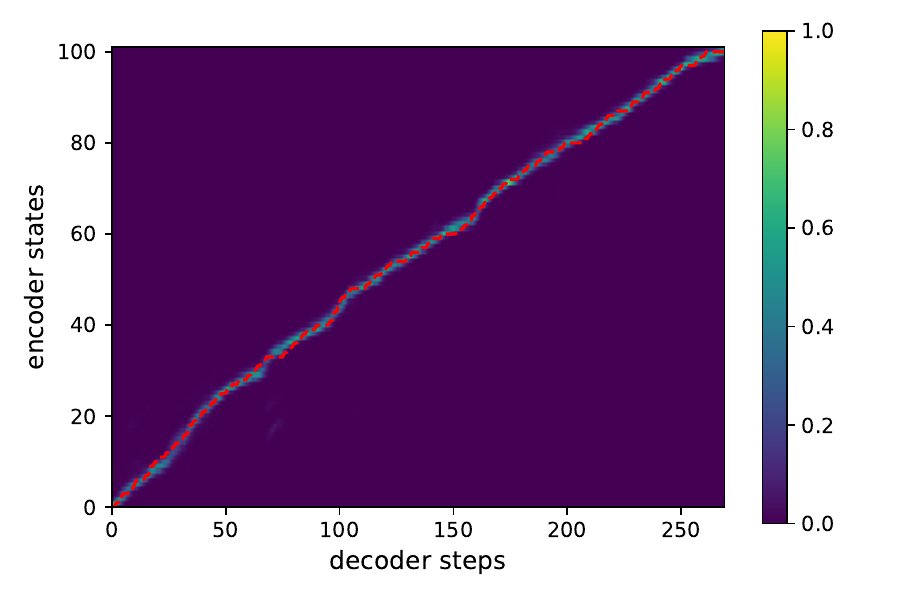}}
\caption{Visualization of the attention alignment and the DTW path of an utterance pair in the validation set. The heat map shows the alignment probabilities
calculated by the attention module in our seq2seq model. The red dashed line shows the alignment path given by DTW, which is downsampled to match the sample rates of encoder states and decoder time steps.}
\label{fig3}
\end{figure}

\subsection{Comparison between different decoder loss functions}
\label{subsec:losseva}

As introduced in Section \ref{subsec:lossfunction}, either MSE or GMM-ML criterion was applied to define the loss function $L_{dec}$ of the decoder output in our implementation.
We evaluated the objective performance of these loss functions by experiments on both female-to-male and male-to-female conversions using the Mandarin dataset.

The Mel-cepstral distortion (MCD) and root mean square error (RMSE) of $F_0$ on validation set were adopted as metrics.
Because Mel-spectrograms were adopted as acoustic features, it's not straightforward to extract $F_0$ and MCCs features from the converted acoustic features.
Therefore, $F_0$ and 25-dimensional MCCs features were extracted by STRAIGHT from the reconstructed waveforms for evaluation.
Then the extracted features were aligned to those of the reference utterances in the validation set in order to compute MCD and $F_0$ RMSE values.
The $F_0$ RMSE was calculated only using the frames which were both voiced in the converted and reference utterances.

\tablename~\ref{tab2} summarizes the objective evaluation results on validation set.
From the table, we can see that the model using the GMM-ML criterion with 2 mixture components achieves the best performance on validation set among all settings except the MCD of male-to-female conversion.
A further examination shows that using the GMM-ML criterion with  mixture components more than 2 may lead to the instability of attention alignment.
Some cases of attention failures, such as getting stuck in one frame, can be observed for MX6 and MX8. We tried to re-optimize the weighting factors in Eq.~(\ref{eq15}) for the MX6 and the MX8. The experimental results showed that changing the coefficients for models with more mixtures could slightly improve the alignment quality while the overall performances of the models were still worse than the MX2 model. One possible reason is that larger mixture numbers may increase the number of parameters and the difficulty of model training.
Thus, the GMM-ML criterion with 2 mixtures was adopted for  $L_{dec}$ in  following experiments.

The SCENT network models pairs of source and target utterance directly. During training, alignments of utterance pairs are learned by attention module implicitly.
An example of the alignment between an utterance pair using the SCENT model is shown in  \figurename~\ref{fig3},
where each column denotes the attention probabilities corresponding to different encoder states for one decoder step.
The DTW algorithm was also conducted based on the input and output Mel-spectrogram sequences and the resulting path was plotted as the red dashed line for comparison.
From this figure, we can see that these two alignments matched well.
Comparing with the DTW path which denotes hard and deterministic alignment,  the attention alignment is soft and changes smoothly along consecutive decoder time steps.

\subsection{Comparison between baseline and proposed methods}
\label{subsec:compari}

\subsubsection{Objective evaluation}

Objective evaluations were first carried out to compare the MCD and $F_0$ RMSE performance of our proposed method and the baseline methods introduced in Section \ref{expcond}, including JD-GMM, DNN, bn-DNN and VCC2018.
In order to compensate the duration differences between source and target speakers,
we also tried to linearly interpolate the source feature sequences before sending them into the conversion models according to the average ratio between the training set durations of the two speakers.
We only interpolated the static part of the source features and the dynamic features were recalculated based on the interpolated static features.
This led to four additional methods, named i-JD-GMM, i-DNN, i-bn-DNN and i-VCC2018, in our evaluations.
The MCDs and $F_0$ RMSEs were calculated following the way introduced in Section \ref{subsec:losseva}.
For fair comparison,  $F_0$ and MCCs  were re-extracted by STRAIGHT from the converted waveforms for all methods when computing MCDs and $F_0$ RMSEs.

\begin{table}[!t]
\renewcommand\arraystretch{1.2}
\caption{Objective evaluation results of baseline and proposed methods on test set of Mandarin dataset.}
\label{tab3}
\begin{tabular}{p{40pt}|p{35pt}<{\centering}|p{35pt}<{\centering}|p{35pt}<{\centering}|p{35pt}<{\centering}}
\toprule
\multirow{3}*{Methods}&
\multicolumn{2}{c|}{Female-to-Male}&
\multicolumn{2}{c}{Male-to-Female}\\
\cline{2-5}
&MCD  &$F_0$ RMSE &MCD &$F_0$ RMSE \\
&(dB)  &(Hz) &(dB) &(Hz) \\
\midrule
JD-GMM &3.892 &55.241 &4.307 &46.625\\
\hline
i-JD-GMM &3.936 &55.939 &4.328 &48.286\\
\hline
DNN &3.688 &44.087 &4.335 &39.190\\
\hline
i-DNN &3.750 &44.268 &4.245 &39.877\\
\hline
bn-DNN &3.618 &42.385 &4.078 &35.883\\
\hline
i-bn-DNN &3.725 &42.961 &4.088 &35.019\\
\hline
VCC2018 &3.802 &56.874 &4.210 &39.196 \\
\hline
i-VCC2018 &3.854 &53.350 &4.225 &41.257 \\
\hline
Proposed &\textbf{3.556} &\textbf{41.748} &\textbf{3.802} &\textbf{33.374} \\
\bottomrule
\multicolumn{5}{p{230pt}}{``i'' represents the interpolation of source features for duration compensation.
``bn'' denotes appending bottleneck features as input.}
\end{tabular}
\end{table}

\begin{table}[!t]
\renewcommand\arraystretch{1.2}
\caption{Objective evaluation results of baseline and proposed methods on test set of English CMU ARCTIC dataset.}
\label{tab:objARC}
\begin{tabular}{p{40pt}|p{35pt}<{\centering}|p{35pt}<{\centering}|p{35pt}<{\centering}|p{35pt}<{\centering}}
\toprule
\multirow{3}*{Methods}&
\multicolumn{2}{c|}{Female-to-Male}&
\multicolumn{2}{c}{Male-to-Female}\\
\cline{2-5}
&MCD  &$F_0$ RMSE &MCD &$F_0$ RMSE \\
&(dB)  &(Hz) &(dB) &(Hz) \\
\midrule
JD-GMM &3.176 &16.473 &3.278 &16.418\\
\hline
i-JD-GMM &3.187 &14.834 &3.274 &16.343\\
\hline
DNN &3.200 &13.998 &3.270 &14.118\\
\hline
i-DNN &3.271 &14.531 &3.296 &14.050\\
\hline
bn-DNN &3.167 &12.675 &\textbf{3.100} &13.070\\
\hline
i-bn-DNN &\textbf{3.141} &11.969 &3.182 &13.098\\
\hline
VCC2018 &3.384 &11.116 &3.668 &13.707 \\
\hline
i-VCC2018 &3.354 &11.455 &3.663 &12.631 \\
\hline
Proposed &3.212 &\textbf{9.899} &3.383 &\textbf{11.704} \\
\bottomrule
\multicolumn{5}{p{230pt}}{``i'' represents the interpolation of source features for duration compensation.
``bn'' denotes appending bottleneck features as input.}
\end{tabular}
\end{table}

The proposed and baseline methods were evaluated on both the Mandarin dataset and the English CMU ARCTIC dataset.
When using the English CMU ARCTIC dataset, the same procedure of tuning the decoder output layer as described in Section~\ref{subsec:losseva} was conducted and the GMM output layer with 2 mixtures was also chosen.

\tablename~\ref{tab3} shows the objective evaluation results of baseline and proposed methods on test set of the Mandarin dataset.
We can see that the MCDs and $F_0$ RMSEs of baseline methods with interpolation were close to or slightly worse than those without interpolation.
Appending bottleneck features as inputs was beneficial for improving the objective performance of the DNN-based method.
Our proposed method outperformed all baseline methods, which obtained the lowest MCD and $F_0$ RMSE.

\tablename~\ref{tab:objARC} shows the results evaluated on the English CMU ARCTIC dataset.
The proposed method achieved best performance on $F_0$ RMSE, while its performance on MCD was not as good as some baseline methods.
Considering that the MCD measurement may be inconsistent with human perception \cite{Toda2007Voice,Chen2014Voice,6542729}, some subjective evaluations were further conducted and
will be introduced later.

\begin{figure}
\begin{minipage}{0.4\textwidth}
\centerline{\includegraphics[width=180pt]{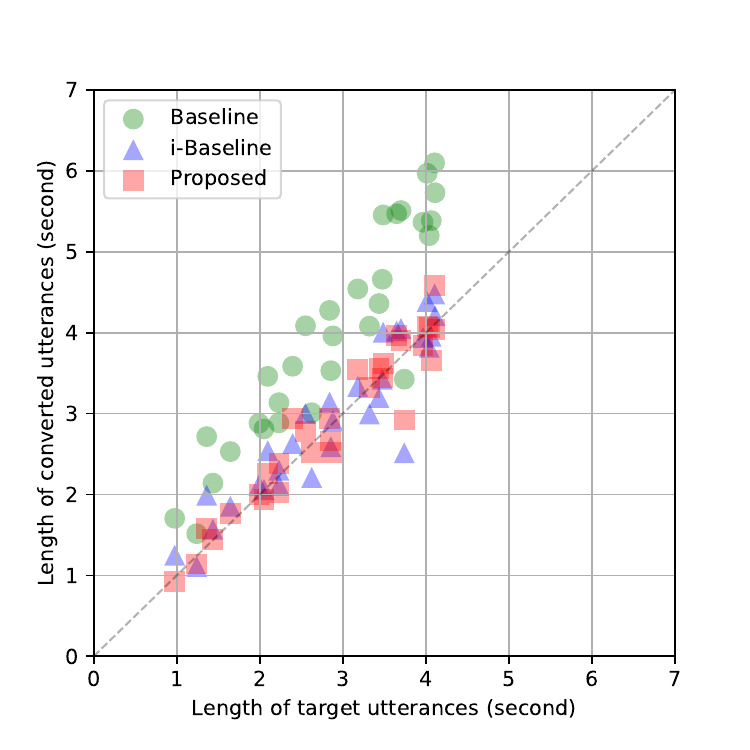}}
\caption{The scatter diagram of the durations of test utterances for female-to-male conversion using the Mandarin dataset.}
\label{fig4}
\end{minipage}
\begin{minipage}{0.4\textwidth}
\centerline{\includegraphics[width=180pt]{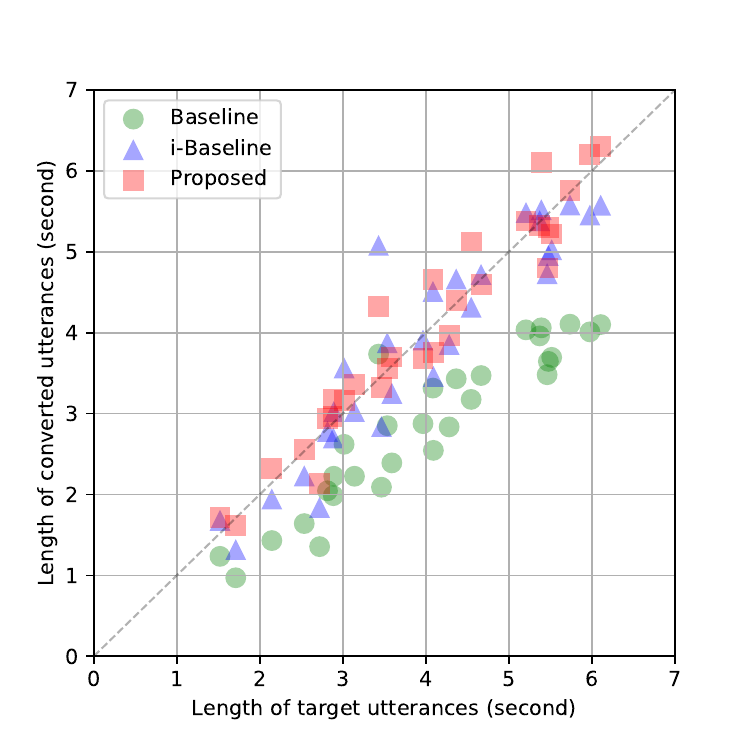}}
\caption{The scatter diagram of the durations of test utterances for male-to-female conversion using the Mandarin dataset.}
\label{fig5}
\end{minipage}
\end{figure}

One advantage of our proposed method is that it can convert the duration of source speech using an unified acoustic model.
In order to investigate the performance of duration conversion,
the scatter diagrams of test utterance durations are drawn in \figurename~\ref{fig4} and \figurename~\ref{fig5} for female-to-male and male-to-female conversions using the Mandarin dataset.
For each test utterance, the durations of  speech converted using different baseline methods were the same, i.e., the duration of the source speech.
For the baseline methods with source feature interpolation, the same global interpolation ratio was shared by all baseline methods.
Therefore, ``i-Baseline'' and ``Baseline'' in these two figures stand for all baseline methods with and without interpolation respectively.

\begin{table}
  \centering
   \caption{The average absolute differences between the durations of the converted and target utterances (DDUR) on test set.}
   \label{tab4}
  \begin{tabular}{p{40pt}<{\centering}|p{43pt}<{\centering}|p{43pt}<{\centering}|p{43pt}<{\centering}}
  \toprule
\multirow{2}*{\tabincell{c}{Conversion \\ Pairs}} & Baseline & i-Baseline & Proposed \\
&   (second) &(second) &(second) \\
  \midrule
   F-M (MA)        & 1.147 & 0.276   &\textbf{0.194} \\
   \hline
   M-F (MA)       & 1.157 & 0.380   &\textbf{0.260} \\
  \hline
   F-M (EN)      &0.560  & 0.282   &\textbf{0.227} \\
   \hline
   M-F (EN)        &0.556  & 0.240   &\textbf{0.147} \\
  \bottomrule
  \multicolumn{4}{p{210pt}}{``F-M'' and ``M-F'' represent female-to-male and male-to-female conversions. ``MA'' and ``EN'' represent the Mandarin and English dataset respectively.}
  \end{tabular}
\end{table}

\begin{figure*}[!t]
\centering
\centerline{\includegraphics[width=0.9\textwidth]{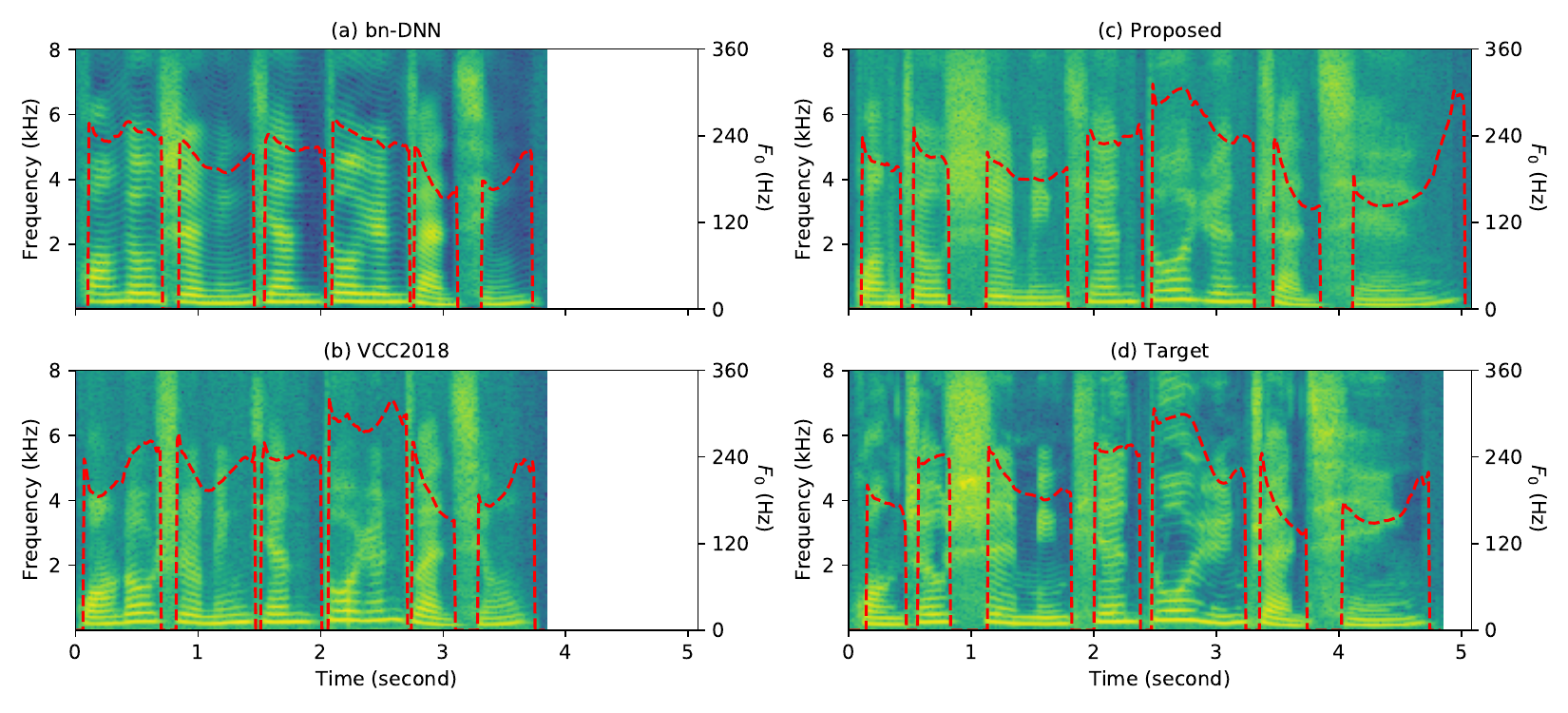}}
\centering
\caption{The $F_0$ contours and spectrograms of one test utterance converted using different methods and the natural target reference.
The red dashed lines are $F_0$ contours extracted by STRAIGHT from the converted waveforms.
}
\label{fig6}
\end{figure*}

From these figures, we can see that the male speaker had higher speaking rate and shorter utterance durations than the female speaker in the Mandarin dataset. The simple linear interpolation made the length of converted speech closer to the target.

Furthermore, the average absolute differences between the durations of the converted and target utterances (DDUR) are calculated using both Mandarin and English datasets and are presented in \tablename~\ref{tab4}.
Results show that our proposed method can generate speech with lower duration errors than the baseline methods without duration modification or with global speaking rate compensation.

\figurename~\ref{fig6} plots the $F_0$ contours and spectrograms of one test utterance converted using different methods and the natural target reference in the Mandarin dataset.
From this figure, we can see that our proposed method can generate speech with more similar $F_0$ contours to the natural reference than the other two baseline methods.
Furthermore, our proposed method can also modify the duration of source speech towards the natural reference appropriately as shown in this figure.

\subsubsection{Subjective evaluation}
\label{subsubsec:subjective}

Subjective evaluations were conducted to compare the performance of our proposed method with the baseline methods in terms of the naturalness and  similarity of converted speech.
In this evaluation, twenty utterances in the test set were randomly selected and converted using our proposed method and three baseline methods,
including i-JD-GMM, i-bn-DNN, and i-VCC2018.

For the experiments conducted on the Mandarin dataset, ten native listeners participated in the evaluation.
For the experiments conducted on the English CMU ARCTIC dataset, evaluations were conducted on the Amazon Mechanical Turk~\footnote{\url{https://www.mturk.com}} (AMT), a platform designed to facilitate crowdsourcing. At least twenty native English listeners took part in the evaluation.
In both evaluations, the listeners were asked to use headphones and the samples were shown to them in random order.
The listeners were asked to give a 5-scale opinion score (5: excellent, 4: good, 3: fair, 2: poor, 1: bad) on both similarity and naturalness for each converted utterance.

\begin{table}
\renewcommand\arraystretch{1.2}
  \centering
   \caption{Mean opinion scores (MOS) with $95\%$ confidence intervals on naturalness and similarity of baseline and proposed methods.}
   \label{tab:mos}
   \resizebox{\linewidth}{!}{
  \begin{tabular}{l|c|c|c|c|c}
  \toprule
  \multicolumn{2}{c|}{\tabincell{c}{Conversion\\Pairs}}&
  i-JD-GMM & i-bn-DNN & i-VCC2018 & Proposed \\
  \midrule
 \multirow{2}*{F-M (MA)} &
 N   & $2.08\pm0.16$  &$2.09\pm0.12$   &$3.29\pm0.10$   &$\textbf{3.70}\pm0.09$ \\
 \cline{2-6}
 & S & $1.86\pm0.13$  &$1.97\pm0.11$   &$2.55\pm0.11$   &$\textbf{3.66}\pm0.09$ \\
 \hline
 \multirow{2}*{M-F (MA)} &
 N   & $1.62\pm0.11$  &$1.78\pm0.12$   &$3.37\pm0.10$   &$\textbf{3.68}\pm0.11$ \\
  \cline{2-6}
 & S & $1.55\pm0.09$  &$1.82\pm0.11$   &$2.29\pm0.11$   &$\textbf{3.80}\pm0.09$ \\
 \hline
 \multirow{2}*{F-M (EN)} &
 N   & $2.90\pm0.13$  &$2.97\pm0.13$   &$3.70\pm0.10$   &$\textbf{3.93}\pm0.10$ \\
 \cline{2-6}
 & S & $3.03\pm0.12$  &$3.11\pm0.11$   &$3.84\pm0.09$   &$\textbf{4.10}\pm0.08$ \\
 \hline
 \multirow{2}*{M-F (EN)} &
 N   & $2.30\pm0.12$  &$2.14\pm0.11$   &$3.72\pm0.10$   &$\textbf{4.10}\pm0.09$ \\
  \cline{2-6}
 & S & $2.58\pm0.11$  &$2.49\pm0.11$   &$3.70\pm0.10$   &$\textbf{4.05}\pm0.09$ \\
  \bottomrule
  \multicolumn{6}{p{245pt}}{``F-M'' and ``M-F'' represent female-to-male and male-to-female conversions respectively. ``MA'' and ``EN'' represent the Mandarin and English dataset respectively. ``N'' and ``S'' denote naturalness and similarity.}
  \end{tabular}}
\end{table}

The results of the subjective evaluations are presented in \tablename~\ref{tab:mos}.
From the table, we can see that the i-bn-DNN method achieved  similar naturalness and similarity to the i-JD-GMM method.
This is consistent with previous studies on DNN-based voice conversion methods \cite{Desai2009voice,Desai2010Spectral,Laskar2012Comparing}.
It should be noticed  that the i-bn-DNN method accepted additional bottleneck features as inputs, which may benefit the performance of this method.
Compared with the i-bn-DNN method, the i-VCC2018 method did not use acoustic features as inputs.
However, this method achieved the best performance among the three baseline methods, especially on the naturalness of converted speech.
One important reason is that the i-VCC2018 method adopted WaveNet vocoder instead of conventional STRAIGHT vocoder to reconstruct speech waveforms from the converted acoustic features.

Our proposed method outperformed the i-VCC2018 method on both naturalness and similarity, also on both Mandarin and English datasets. These experimental results proved the effectiveness of our proposed method and the improvement brought by our proposed method was not limited to a specific language.
One possible reason is that at the conversion stage of  the i-VCC2018 method, bottleneck features extracted from source speech were fed to the acoustic predictor. While the model was trained with the bottleneck features of the target speaker as inputs \cite{Liu2018}.
This inconsistency may degrade the similarity of converted speech.
Another reason can be attributed to the duration conversion ability of our proposed method as introduced in the objective evaluations.
Therefore, the prosody similarity and naturalness of our proposed method  were better than simply adjusting speaking rate globally.

\begin{figure*}[!t]
\centering
\centerline{\includegraphics[width=0.9\textwidth]{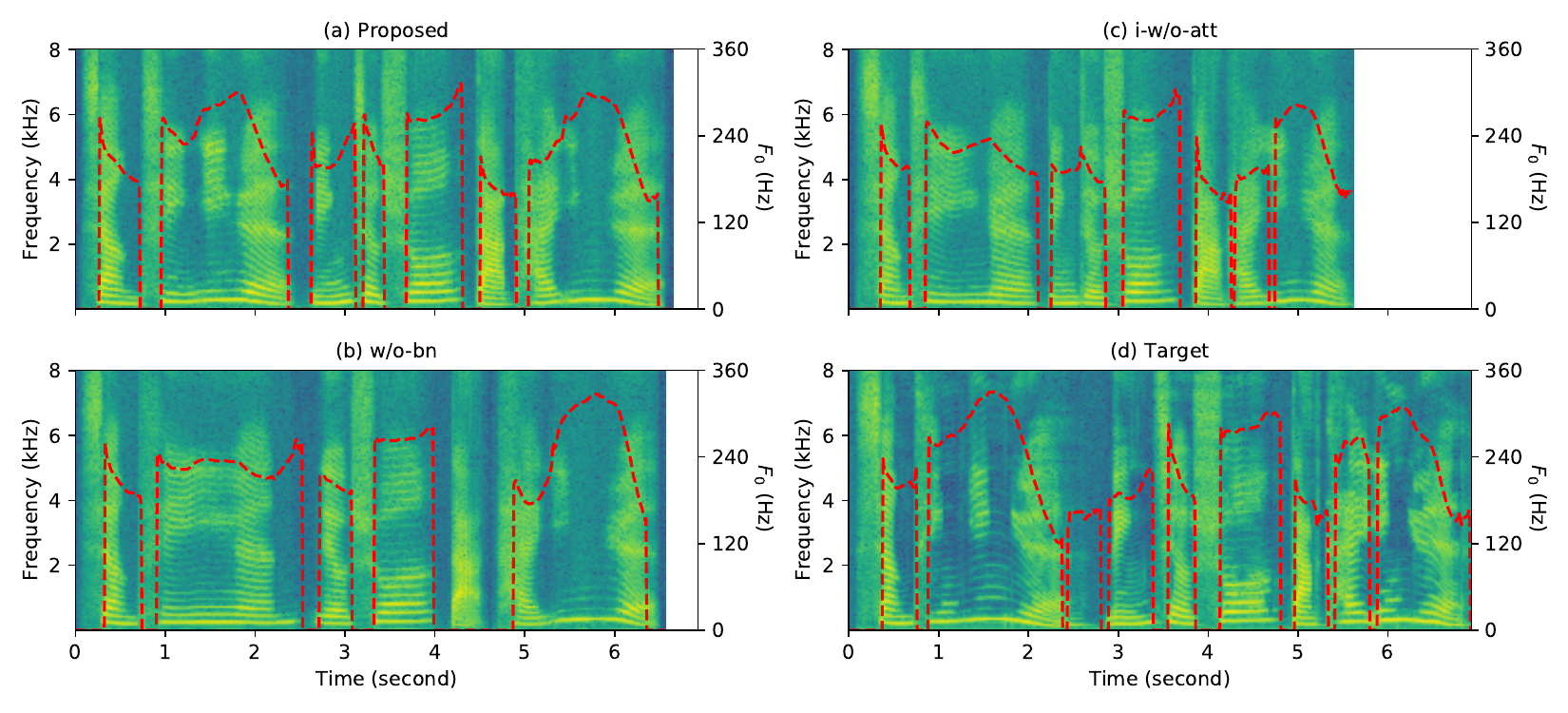}}
\centering
\caption{The $F_0$ contours and spectrograms of one test utterance converted using different methods and the natural target reference. ``w/o-bn'' and ``i-w/o-att'' represent the proposed models without bottleneck features and without attention module but adjusting speaking rate globally by interpolation respectively.
The red dashed lines are $F_0$ contours extracted by STRAIGHT from the converted waveforms.
}
\label{fig9}
\end{figure*}

\subsection{Ablation tests}
\label{subsec:ablation}

In order to further analyze the effectiveness of some key components in our model, ablation tests on model inputs, attention module and  location code  were conducted. In this subsection, only the Mandarin dataset was adopted for evaluation.

\subsubsection{Mel-spectrograms and bottleneck features}

\begin{table}[!t]
\renewcommand\arraystretch{1.2}
\caption{Objective evaluation results of proposed methods without using Mel-spectrograms and without using bottleneck features as inputs.}
\label{tab5}
\begin{tabular}{p{30pt}|p{35pt}<{\centering}|p{35pt}<{\centering}|p{35pt}<{\centering}|p{35pt}<{\centering}}
\toprule
\multirow{3}*{Methods}&
\multicolumn{2}{c|}{Female-to-Male}&
\multicolumn{2}{c}{Male-to-Female}\\
\cline{2-5}
&MCD  &$F_0$ RMSE  &MCD  &$F_0$ RMSE \\
&(dB) &(Hz) &(dB) &(Hz)\\
\midrule
Proposed &\textbf{3.556} &\textbf{41.748} &\textbf{3.802} &\textbf{33.374}\\
\hline
w/o-Mel &3.623 &43.443 &3.803 &35.463 \\
\hline
w/o-bn &3.624 &48.550 &4.000 &40.183 \\
\bottomrule
\multicolumn{5}{p{220pt}}{``w/o-Mel'' and ``w/o-bn''  represent the models without using Mel-spectrograms
and without using bottleneck features as inputs respectively.}
\end{tabular}
\end{table}

In order to investigate the necessity of using Mel-spectrograms and bottleneck features, we removed each one of them and built SCENT models  utilizing only source bottleneck features or Mel-spectrograms as inputs respectively.
Objective evaluation results of MCD and $F_0$ RMSE on test set are presented in \tablename~\ref{tab5}.

From this table, we can see that Mel-spectrograms are beneficial for the model to achieve more accurate prediction of acoustic features.
It also can be found that removing bottleneck features led to higher $F_0$ RMSE and MCD on test set, and its degradation on $F_0$ RMSE was more serious than removing Mel-spectrograms.
Listening to the converted audio samples without using bottleneck features, we found they suffered from serious mispronunciation problem.
The bottleneck features extracted by an ASR model contain high-level and linguistic-related information.
The experimental results indicate that they were essential for achieving stable voice conversion results in our proposed method.

$F_0$ contours and spectrograms of one test utterance converted by the proposed method and the proposed method without bottleneck features are presented in \figurename~\ref{fig9} (a) and \figurename~\ref{fig9} (b) respectively. Compared to the method without using bottleneck features, the $F_0$ contour of the utterance converted by our proposed method is more similar to that of the natural reference in \figurename~\ref{fig9} (d). Also, a significant spectral distortion can be observed at the $1\sim2s$ interval of the spectrogram generated by the  ``w/o-bn'' method.

\subsubsection{Attention module}

\begin{table}[!t]
\renewcommand\arraystretch{1.2}
\caption{Objective evaluation results of proposed methods with and without the attention module.}
\label{tab6}
\begin{tabular}{p{40pt}|p{35pt}<{\centering}|p{35pt}<{\centering}|p{35pt}<{\centering}|p{35pt}<{\centering}}
\toprule
\multirow{3}*{Methods}&
\multicolumn{2}{c|}{Female-to-Male}&
\multicolumn{2}{c}{Male-to-Female}\\
\cline{2-5}
&MCD &$F_0$ RMSE &MCD &$F_0$ RMSE \\
&(dB) &(Hz) &(dB) &(Hz)\\
\midrule
Proposed &\textbf{3.556} &\textbf{41.748} &\textbf{3.802} &\textbf{33.374} \\
\hline
w/o-att &3.635 &47.620 &3.969 &37.948 \\
\hline
i-w/o-att &3.770 &50.310 &3.914 &37.034 \\
\bottomrule
\multicolumn{5}{p{230pt}}{``w/o-att'' and ``i-w/o-att'' represent models without attention module and without attention module but adjusting speaking rate globally by interpolation respectively.}
\end{tabular}
\end{table}

The attention module in a SCENT model helps to achieve the alignment between input and output feature sequences at the training stage and to predict target durations at the conversion stage.
In order to investigate how the attention module contributed to the overall performance of our proposed method,
we modified the SCENT model to a frame-by-frame transformation model without attention mechanism for comparison.
Once the attention module was removed, the  LSTM layer with attention in the decoder became a plain uni-directional LSTM.
In order to get frame aligned sequence pairs for model training, the input sequences were wrapped towards the target ones using DTW algorithm and MCCs features.
The other parts of the  SCENT model were kept unchanged.

Our experiments compared three methods, including the proposed method, the proposed method without attention (w/o-att) and the proposed method without attention but using source interpolation at conversion time (i-w/o-att).
\tablename~\ref{tab6} shows the MCDs and $F_0$ RMSEs of these three methods.
We can see that the prediction errors increased in the absence of the attention module.
$F_0$ contours and spectrograms of one test utterance converted by proposed method and ``i-w/o-att'' method are presented in Fig.~\ref{fig9} (a) and Fig.~\ref{fig9} (c) respectively. This figure again shows the effectiveness of the attention module for generating speech with duration and $F_0$ contour closer to that of the natural target speech.

\begin{table}[!t]
\renewcommand\arraystretch{1.2}
  \centering
  \caption{The results of preference tests on similarity among proposed methods with and without the attention module.}
  \resizebox{\linewidth}{!}{
  \begin{tabular}{l|p{27pt}<{\centering}|p{28pt}<{\centering}|p{28pt}<{\centering}|p{28pt}<{\centering}|c}
  \toprule

  &w/o-att &i-w/o-att &Proposed & N/P &\multirow{2}*{$p$}\\
  & (\%) &(\%) & (\%) &(\%) & \\
  \midrule
  \multirow{2}*{F-M} &
  33.0 & \textbf{58.5} &  -    & 8.5  & $1.31\times10^{-4}$ \\
  \cline{2-6}
  &   -  & 21.0 & \textbf{67.5} & 11.5 & $<1\times10^{-9}$   \\
  \hline
  \multirow{2}*{M-F} &
  17.5 & \textbf{76.0}   &  -    & 6.5  & $<1\times10^{-9}$  \\
  \cline{2-6}
  & -     & 24.0 & \textbf{66.5} & 9.5  & $<1\times10^{-9}$  \\
  \bottomrule
  \multicolumn{6}{p{230pt}}{ ``$p$'' represents $p$ value of $t$-test.
  ``N/P'' denotes no preference. ``F-M'' and ``M-F'' represent female-to-male and male-to-female conversions respectively.}\label{tab7}
  \end{tabular}}
\end{table}

Furthermore, a group of preference tests were conducted  to compare the subjective performance of these three methods.
Because the duration conversion achieved by the attention module significantly affected on the similarity,
the preference tests focused on the similarity aspect of converted speech.
Ten native listeners were involved in evaluation and the experimental results are presented in \tablename~\ref{tab7}.
This table shows that the strategy of global speaking rate adjustment by source interpolation can improve the similarity of converted speech in both conversion pairs.
The proposed method with attention module outperformed the method without attention but using source interpolation.
These results further confirmed the effectiveness of the attention module.

\subsubsection{Location code}

\begin{table}[!t]
\renewcommand\arraystretch{1.2}
\centering
\caption{Objective evaluation results of proposed methods with and without the location code.}
\label{tab:loc}
\begin{tabular}{p{25pt}<{\centering}|p{35pt}<{\centering}|p{35pt}<{\centering}|p{35pt}<{\centering}|p{35pt}<{\centering}}
\toprule
\multicolumn{2}{c|}{\multirow{2}*{Methods}}&
 MCD&  $F_0$ RMSE&  DDUR \\

 \multicolumn{2}{c|}{~} &(dB) & (Hz)  &(second)\\
\midrule
\multirow{2}*{F-M} &
Proposed & \textbf{3.556} & \textbf{41.748} & \textbf{0.194} \\
\cline{2-5}
& w/o-locc  & 3.590 & 41.783 & 0.205  \\
\hline
\multirow{2}*{M-F} &
Proposed & \textbf{3.802} & \textbf{33.374} & \textbf{0.260} \\
\cline{2-5}
& w/o-locc &  3.822 & 35.561 & 0.307 \\
\bottomrule
\multicolumn{5}{p{220pt}}{``DDUR'' represents average absolute difference
between the durations of the converted and target utterances. ``w/o-locc'' represents models without location code. ``F-M'' and ``M-F'' represent female-to-male and male-to-female conversions respectively.}
\end{tabular}
\end{table}

Ablation tests were conducted for investigating how the location code affected the performance of the model.
In the experiments, the location code was removed and the models were built in the same conditions.
MCD, $F_0$ RMSE and DDUR were calculated and are presented in \tablename~\ref{tab:loc}. A slight raise of MCD and $F_0$ RMSE after removing the location code can be observed from this table.
Furthermore, the DDURs in female-to-male and male-to-female conversions increased by 5.4\% and 15.3\% respectively. These experimental results demonstrated the positive effects of using the location code.

\subsection{Discussions}
\label{sec:discussion}
As discussed in Section \ref{sec:relatedwork}, directly implementing seq2seq models at utterance level is difficult for the voice conversion task.
The input and output sequences in voice conversion are composed of frame-level features and are relatively long thus it is a challenge for the attention mechanism to search for the correct hidden entries to pay attention to.
Once there are abnormal skips or repetitions in the sequence of attention probabilities, mistakes of converted speech may occur.

These difficulties are considered when designing the \net model. In order to improve attention stability, the techniques of forward attention and adding location features are used when calculating attention probabilities.
The bottleneck features can also provide linguistic-related information to help the attention-based alignment between input and output feature sequences.
However, errors still can not be completely avoided in the converted speech.
Additional 100 non-parallel utterances of both speakers in the Mandarin dataset, which were out of the dataset used for previous experiments, were adopted for error analysis.
The utterances of the male speaker contained 2747 phonemes, while the utterances of the female speaker had 2538 phonemes.
We conducted male-to-female and female-to-male conversions for these utterances and identified different categories of conversion errors subjectively.
In male-to-female conversion,  there were 1 skipping phoneme error,  2 completion prediction errors, 34 phoneme pronunciation errors, 31 tone defects and 10 phoneme quality defects.
In female-to-male conversion, there were 19 phoneme pronunciation errors, 20 tone defects and 17 phoneme quality defects.

Several reasons may lead to these errors. First, the proposed model contains about 7.5 M trainable parameters thus is complex and needs to be trained in a data-driven way.
Therefore, the insufficiency of training data may cause the model's lack of  generalization ability when dealing with unseen utterances.
Also, the extracted bottleneck features may also be misleading due to the accuracy limitation of the ASR model.
To further reduce conversion errors and to produce more reliable conversion results using seq2seq models will be an important task of our future work.

\section{Conclusion}
\label{sec:conclusion}
This paper presents SCENT, a sequence-to-sequence neural network, for  acoustic modeling in voice conversion.
Mel-spectrograms are used as acoustic features. Bottleneck features extracted by an ASR model are taken as additional linguistic-related descriptions and are concatenated with the source acoustic features as network inputs. Taking advantage of the attention mechanism, the SCENT model does not rely on the preprocessing of DTW alignment and the duration conversion can be achieved simultaneously.
Finally, the converted acoustic features are passed through a WaveNet vocoder to reconstruct speech waveforms.
Objective and subjective experimental results demonstrated the superiority of our proposed method compared with baseline methods, especially in durational aspect.
Ablation tests further proved the benefits of inputting Mel-spectrograms and the necessity of bottleneck features. The importance of the attention module and the positive effect of the location code were also proved in our ablation studies.
To investigate the influence of training set size on the performance of our proposed method and to reduce conversion errors by improving attention calculation will be our work in the future.

\bibliographystyle{IEEEtran}
\bibliography{references}

\end{document}